\documentclass[aps]{revtex4}
\usepackage{graphicx}%

\begin{document}
\title{Phase transitions in one dimension: 
are they {\em all} driven by domain walls? }
\author{Nikos Theodorakopoulos$^{1,2}$} 
\affiliation{
$^{1}$Theoretical and Physical Chemistry Institute, National Hellenic Research Foundation,
Vasileos Constantinou 48, 11635 Athens, Greece\\
$^{2}$Fachbereich Physik, Universit\"at Konstanz, 78457 Konstanz, Germany\\
}
\date{\today}
\begin{abstract}
Two known distinct examples of one-dimensional systems which are known to 
exhibit a phase transition are critically examined:
(A) a lattice model with harmonic nearest-neighbor elastic
interactions and an on-site Morse potential, 
and (B) the ferromagnetic, spin $1/2$ Ising
model with long-range pair interactions varying as the inverse square of the
distance between pairs.  In both cases it can be shown that the 
domain wall configurations become entropically stable at, or very near, the critical temperature.
This might provide a \lq\lq positive\rq\rq\- criterion for the occurrence of a phase transition
in one-dimensional systems. 
 
\end{abstract}
\maketitle

\section{Introduction}
Phase transitions do not generally occur in one-dimensional systems. Their absence has been
codified under a variety of conditions. Van Hove \cite{vanHove} showed that, for particle systems with pair interactions 
of sufficiently short range, the free energy does not have any singularities. Landau \cite{Landau} argued that
a bistable system with finite interface energy  will achieve an entropic gain by breaking
up into a macroscopic number of domains, making 
macroscopic phase coexistence at any positive temperature impossible. His argument explains the absence of a phase transition in a 
large class of one-dimensional systems, which includes the short-range $S=1/2$ Ising model, the $\phi ^{4}$ model, etc. 

How does theory cope with the exceptions from the broad rule? Proofs of the van Hove type are in essence  
proofs of analyticity. In principle they can be extended, refined and thus provide sharper
criteria for the {\it absence} of a phase transition\cite{Cuesta}.  Nonetheless, it is 
not straightforward to 
make a positive statement about a system 
where a phase transition actually occurs.   On the other  hand, Landau's approach turns out to be more 
potent, because it can be \lq\lq inverted\rq\rq\-  to describe the exact pathology which is necessary to produce macroscopic phase 
coexistence. Instead of setting out to prove that a finite interface energy will lead to a macroscopic number of 
domains  because of the entropic gain invoved, 
it is possible to look at a single, possibly pathological  interface, or domain wall (DW),
and examine under which conditions the entropic gain might lead to its spontaneous formation. The question at hand
is whether the conditions for spontaneous formation of a single interface are the same as those for a thermodynamic 
instability of the low-temperature phase, or, equivalently, for
macroscopic phase coexistence.  If the conditions are the same, i.e. if (i) spontaneous DW formation occurs at some
temperature $T^{*}$, (ii) a macroscopic instability / phase transition is known by independent means
to occur at a temperature $T_{c}$, and (iii) $T^{*}=T_{c}$, then it is not unreasonable to argue 
that the transition is driven by the formation of the DW. 

In this paper I will 
review how this can be done in two representative cases: (A) models of one-dimensional 
thermodynamic instabilities (interfacial wetting, DNA denaturation),  (B) the $S=1/2$ ferromagnetic Ising model with 
inverse-square interaction. Case (A) will be illustrated by the Peyrard-Bishop \cite{PB} Hamiltonian (continuum and discrete)
 - although this should be regarded as representative of a much broader class of on-site potentials with a 
repulsive core, a stable minimum and a flat top.  A proper understanding of how DW formation 
governs one-dimensional phase transitions includes a discussion of symmetric on-site potentials, where the pathology is  
 \lq\lq accidentally\rq\rq\- lifted and a phase transition does not occur. In a sense this represents a special case of
(A) and will be treated accordingly.  Case (B) has been recognized for a long time as a key representative of marginal
pathology; in particular, Thouless \cite{Thou69} has used a Landau-type of argument  to predict the occurrence of
long-range order at low temperatures; his argument however does not lead to an estimate of the critical temperature. It turns out that a 
\lq\lq literal\rq\rq\- application of the single-DW criterion (spontaneous DW formation), 
which has been successful in the case of thermodynamic 
instabilities, yields an estimate of the critical temperature which agrees with state-of-the-art Monte Carlo 
calculations\cite{Luijten01}. This simple result is apparently new; it motivates the question stated in 
the title of the paper. 
In view of the pivotal role played by Serge Aubry in making our community aware of 
the relevance of nonlinear excitations (domain walls, solitons etc.) to 
statics and dynamics of phase transitions \cite{AubI,AubII,KRUMSCHRIEF}, 
I believe this is a most appropriate place to discuss it.
 
The paper is organized as follows: Section II briefly restates the original
Landau argument and a variant which will be useful here. The next two sections deal 
with cases (A) and (B), respectively.  Conclusions are summarized and discussed in the final section.

\section{A variant of the Landau argument}
\subsection{Landau's original argument\cite{Landau}}
Given a bistable system with $N$ sites and 
an interface energy $E$ separating phases $A$ and $B$, the number of interfaces $M$ is
such as to minimize the total free energy
\begin{equation}
F(M) = M E - T S(M)
\label{eq:LanF}
\end{equation}
where 
\begin{equation}
S = \ln\left[ \frac{N!}{M!(N-M)! }\right] \approx \ln \left(  \frac{Ne}{M} \right) 
\label{eq:LanS}
\end{equation}
is the entropy; the second expression in (\ref{eq:LanS}) involves use of Stirling's formula.
Minimization of $F$ with respect to $M$ yields the most probable value of $M$
\begin{equation}
{\bar M} = N e^{-E/T}  \quad.
\label{eq:LanM}
\end{equation}
At any $T>0$ the system breaks up into a macroscopic number of domains; in other words, 
phase coexistence cannot occur at a macroscopic scale.

\subsection{Comments and variations}
An obvious way to circumvent the prohibition (\ref{eq:LanM}) is an interface (domain wall (DW))
energy which grows logarithmically with the system size. If $E=\alpha \ln N$, the number 
of domains is macroscopic only for $T>\alpha $. At $T=\alpha $ macroscopic phase coexistence
can occur. Unfortunately, such a dependence of $E$ on $N$ 
is usually accompanied by edge effects,
i.e. the energy of the DW is no more independent of its position. In order to deal with such cases,
it is more convenient to focus on a single DW, situated at position $j$, and examine its thermal
properties. 
The DW partition function is given by
\begin{equation}
Z_{DW} = \sum_{j=0}^{N-1}e^{-E_{DW}(j)/T} \quad.
\label{eq:Zgen}
\end{equation}
I will make explicit use of this in the context of case (B) below. For the moment, note that
this formulation recovers the original Landau result if 
 $E(j) = E \> \forall j$; in that case  $Z_{DW}=N \exp (-E/T)$, and the DW free energy is
\begin{equation}
F_{DW}= E - T \ln N \quad.
\label{eq:DWfeu}
\end{equation}
It follows directly from (\ref{eq:DWfeu}) that a DW will form spontaneously at a temperature
$T^{*}=E/\ln N$, i.e.,  as long as $E$ is finite, 
no phase coexistence at any nonzero temperature is possible in the thermodynamic
limit. Moreover, if $E=\alpha \ln N$, the DW free energy will vanish - and phase coexistence
can occur - at $T^{*}=\alpha$, just as predicted
by (\ref{eq:LanM}).

The vanishing of the DW free energy at a nonzero temperature 
provides an alternative, \lq\lq positive\rq\rq\- criterion
for phase coexistence. I will use - and test - it in the remainder of this paper.

\section{case A: thermodynamic instabilities}
\subsection{Background: Definitions}
Consider the class of one-dimensional Hamiltonians
\begin{equation}
H  = \sum_{n=0}^{N}\Biggl[     \frac{p_n^2}{2}   + 
  \frac{1}{ 2R} (y_{n+1}-y_{n})^{2} + V(y_n)   \Biggr]
\label{eq:PBHam}
\end{equation}
where $y_n$,  $p_n$ are, respectively, the displacement and momentum of the $n$th particle, 
$R$ is a 
coupling constant  and $V(y)$ is any 
potential with a  repulsive core, a stable minimum and a flat top. A convenient representative of this
class is the Morse potential %
\begin{equation}
V(y)= (1 - e^{-y})^{2}  \quad .
\label{eq:Morse}
\end{equation}
All quantities described above are dimensionless. 
The model defined by Eqs. (\ref{eq:PBHam})-(\ref{eq:Morse})
has been proposed in a variety of physical contexts, including interfacial wetting \cite{KroLip}
and DNA denaturation \cite{PB}. 
\subsection{Background: Thermodynamics}
The 
thermodynamic properties of (\ref{eq:PBHam})
are described by the classical partition function, whose nontrivial, 
configurational part  $Z$
can be expressed in terms of the spectrum 
of the symmetric 
transfer integral (TI) operator
\begin{equation}
\int_{ -\infty }^{ \infty } dy' \> K(y,y')
\phi_{\nu }(y') =e^{-\epsilon _{\nu} /T}  \phi_{ \nu }(y)  \quad,
\label{eq:TI}
\end{equation}
where 
\begin{equation}
K(y,y') = e^{- (y'-y)^{2 }/2R T  }
\> e^{-  \left[V(y)+V(y')  \right]/2 T} \quad.
\label{eq:kernel}
\end{equation}
In the thermodynamic limit $N \to \infty $, 
\begin{equation}
Z   =  \sum_{\nu}^{} e^{-N\epsilon _{\nu }/T} 
\label{eq:Zconf}
\end{equation}
is dominated by the smallest eigenvalue
$\epsilon_0$. The free energy per site is 
\begin{equation}
f= - \frac{T}{N } \ln Z \approx  \epsilon _{0} .
\label{eq:fr}
\end{equation}
\lq\lq Analyticity" proofs examine the spectrum of the TI operator. For example, if the potential $V(y)$ 
is unbounded from above as $y \to \pm \infty$, it can be immediately shown \cite{Parisi} that the operator in (\ref{eq:TI})  
is of the Hilbert-Schmidt type, and therefore the spectrum is nondegenerate at any $T>0$, i.e. there are
no singularities in the free energy. This scenario does not hold for the Morse potential - or any other 
potential of its class (cf. above) - which leads to a non-Hilbert-Schmidt kernel  \cite{Zhang}.  
Finite-size scaling \cite{nth_fss} suggests that a single bound state gradually merges into the continuum, i.e. 
\begin{equation}
\epsilon _{0} (T)  \propto \left[T_c(R) - T\right]^{2}  \quad,
\label{eq:e0T}
\end{equation}
where $T_c(R)$ must, in general, be determined numerically.
If the coupling constant is small, $R\ll 1$, the harmonic coupling in (\ref{eq:PBHam}) is strong. It is possible 
to make a continuum aproximation $y_n \to y(x)$. To leading order, this procedure in effect
approximates the integral equation
(\ref{eq:TI}) by the Schroedinger-like equation
\begin{equation}
\left[  -\frac{RT^{2}}{2} \frac{d^{2}}{ dy^{2}} + V(y) \right] \phi_{ \nu }(y)  =  
\epsilon _{\nu }\phi_{ \nu }(y) \quad.
\label{eq:TISchr}
\end{equation}
In the case of (\ref{eq:TISchr}), the above scenario, 
i.e. the quadratic disappearance of the bound state into the continuum, can be verified analytically:
Eq. \ref{eq:e0T} holds exactly, with
\begin{equation}
T_c(R) =  T_{c}^{cont}(R)= 2 \left( \frac{2}{R } \right)^{1/2}    \quad.
\label{eq:TcR}
\end{equation}
Other thermodynamic properties can also be calculated exactly; for example, it is possible
to show \cite{DTP} that the average displacement
\begin{equation}
<y> \approx \frac{T_{c}}{T_{c}-T}  
\label{eq:MorseOP}
\end{equation}
diverges linearly in the neighborhood of $T_c$.
\subsection{The exact DW and its properties}
The equations of motion which correspond to the continuum limit of the 
Hamiltonian (\ref{eq:PBHam}) admit  exact static solutions of the
type \cite{DTP}
\begin{equation}
y(x) = \ln \left[  1  + e ^{\pm (x-x_0) / d}\right]
\label{eq:DWeq}
\end{equation}
where $d=1/(2R)^{1/2}$ and $x_0$ is an arbitrary constant. These solutions have the property  
that they vanish as $x \to \mp \infty$ and become linearly unbounded as $x \to \pm \infty$. 
They can be regarded as interpolations
from the bound to the unbound phase, i.e. they represent {\it bona fide } DWs.  For any
given transverse displacement of the boundary atom $y_{N+1}=Y$, $y_0=0$, the
excess energy of the DW, compared to the minimum of the Morse potential,
is equal to $2 \nu$, where $\nu=Y/(2R)^{1/2}$ 
is the number of unbound sites (not necessarily an integer, but this is not important for the present argument 
in the continuum limit). Their excess entropy, compared to the entropy of the bound chain, 
has been found \cite{DTP} to be $(R/2)^{1/2}\nu$. The DW excess free energy,
\begin{equation}
F_{DW}=\left[ 2 - T \left(\frac{R}{2 } \right)^{1/2} \right] \nu
\label{eq:FMorseDW}
\end{equation}
vanishes at $T^{*}= 2(2/R)^{1/2}$, i.e. at the critical temperature (\ref{eq:TcR}). 
The above argument has been generalized \cite{TPM} for DWs of the highly discrete ($R \gg 1$) version of the Hamiltonian  
(\ref{eq:PBHam}); the numerical agreement between $T^{*}$ and $T_c$ is better than $1\%$.

\subsection{Symmetric on-site potentials}
It is well known that Schroedinger-like equations of the type (\ref{eq:TISchr}) with symmetric on-site potentials which
are bounded from above will support bound states for any value of the control parameter $R$. 
The average
value of the displacement $<y>$ vanishes identically, since the ground state is
of even parity. The system remains bound at all temperatures.

The DW picture provides a straightforward explanation why
such an on-site potential cannot generate a phase transition: in addition to the DW solutions (\ref{eq:DWeq})
interpolating from the stable minimum to $y \to +\infty $, there is a symmetric pair of solutions interpolating
from the stable minimum to $y \to -\infty $.  Thus, although a particular DW solution can become entropically
stable, an open system will always have an equal chance of creating the symmetric DW; the average 
displacement will vanish at all temperatures.
 


\section{case B: The ferromagnetic $S=1/2$ Ising  model with inverse square interactions}
\subsection{Background: Definitions, a glimpse of history \& known estimates of $T_c$}
The one-dimeensional, ferromagnetic $S=1/2$ Ising Hamiltonian with long range interactions is defined by
\begin{equation}
H = -\frac{J}{2 }\sum_{m,n}\frac{\sigma_{m} \sigma_{n} }{ |m-n|^{\lambda }}
\label{eq:lrIsing} 
\end{equation}
where $J>0$, $\sigma _{n}=\pm 1$ is known to have a phase transition if $\lambda \leq 2$; the marginal case
$\lambda =2$ is of particular interest; Thouless \cite{Thou69} predicted the occurrence of long-range order
at low temperatures on the basis of a Landau-like argument.
Anderson and Yuwal showed its relationship to the Kondo Hamiltonian
and estimated $K_{c}= J/T_{c}=0.635$ \cite {AnYu}. In spite of considerable theoretical progress during 
the years that followed \cite{Luijten97},
the exact value of the critical point remains unknown. A state-of-the-art Monte Carlo calculation by 
Luijten \cite{Luijten01} gives an
estimate $K_{c}=0.6552(2)$. Other published 
estimates differ from each other by as much as 30\% \cite{Luijten01, LuijtenRev}.

\subsection{The DW and its energy}
The spin configuration of (\ref{eq:lrIsing}) with minimal energy is $ \sigma _{n} = \mp 1, \forall n$.
A DW positioned at $n=L$ is defined as a spin configuration with 
\begin{equation}
 \sigma _{n} = \left\{ \begin{array}{ll}
    & \mp1 \quad \mbox{if $n\leq L$    } \\
    & \pm1\quad \mbox{if   $n > L$     } 
\end{array}
\right.
\end{equation}
	I will choose the upper sign; the excess energy $E_{DW}(L)$ of the DW (i.e. after subtracting the energy of the ground 
state, cf. above) can easily be calculated as
\begin{equation}
\frac{E_{DW}(L)}{2J } = \Sigma_L (\lambda -1)-L  \Sigma_L (\lambda) +
 \Sigma_{N-L-1} (\lambda - 1) - (N-L) \Sigma_{N-L-1} (\lambda) 
- [ \Sigma_{N-1} (\lambda-1) - N \Sigma_{N-1} (\lambda) ]  \quad, 
\label{eq:DWEner}
\end{equation}
where
\begin{equation}
 \Sigma_n (\lambda) = \sum_{l=1}^{n}\frac{1}{l^{\lambda}} \quad.
\label{eq:sum}
\end{equation}
Eq. (\ref{eq:DWEner}) is valid for $L\leq N/2$; energies at larger values of $L$ can be computed
by using the symmetry $E_{DW}(L)=E_{DW}(N-L)$. 

For $\lambda>2$ it is straightforward to establish that the DW energies are finite and essentially 
$L-$independent - except for a narrow range near the edges, which becomes vanishingly small
in the thermodynamic limit. Eqs. (\ref{eq:DWfeu}) and 
(\ref{eq:LanM}) are both
strictly applicable; either of them can be used to predict the absence
of a phase transition. 

\begin{figure}
\vskip -.5truecm
\includegraphics[width=65mm]{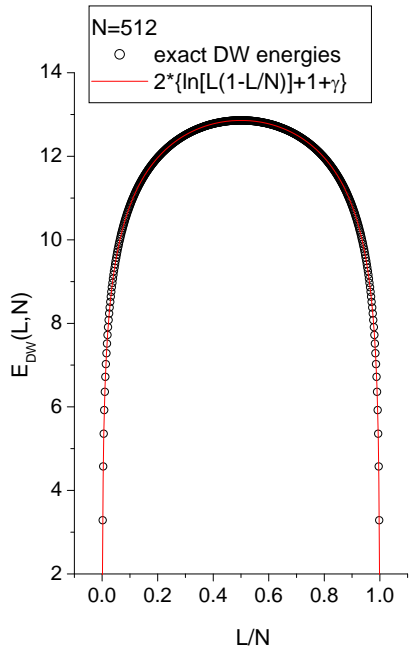}
\vskip -.5truecm
\caption{ 
{\small 
The excess energy of a DW positioned at the $L$th site of a chain with $N=512$, in units of $J$.
The points represent exact values according to (\ref{eq:DWenex}), the continuous line is the approximation of neglecting the contribution
of $f(L)$ (cf. text).
}
}
\label{fig:DWen}
\end{figure}

The marginal case $\lambda=2$ is more interesting. One needs
\begin{eqnarray}
\Sigma_n (2) &= & \sum_{l=1}^{n}\frac{1}{l^{2}} =  \zeta(2) - \psi ' (1+n) \\
\Sigma_n (1) &= & \sum_{l=1}^{n}\frac{1}{l}        = \gamma + \psi  (1+n)\quad,
\end{eqnarray}
where $\gamma$
is the Catalan constant and $\psi$, $\psi'$ the digamma and trigamma
functions, respectively. It is then possible to rewrite the DW energy - now for the special case $\lambda=2$
as 
\begin{equation}
\frac{E_{DW}(L)}{2J}
 = \ln \left [L\left(1-\frac{L}{N}\right)\right] + \gamma + f(L)
\label{eq:DWenex}
\end{equation}
where the last term 
\begin{equation}
f(L) = \psi(1+L) - \ln L  -1 + L\psi' (1+L)
\label{eq:fL}
\end{equation}
is always small. This can be seen in Fig.  \ref{fig:DWen}, where I plot the exact energy and the approximation of entirely neglecting $f$;
$f$ can be shown to be entirely negligible for large $L$ (of order $(1+L)^{-2}$); it is largest at $L=1$, but even then its
value is  $\psi(2)+\psi'(2)-1 = 0.06772$.

\subsection{The DW's entropy: analytical approximation}
Neglecting the contribution of the $f$-term (cf. previous subsection), it is possible to write down an
approximate partition function
\begin{equation}
Z_{DW}^{*} = e^{-2(\gamma+1)K} \sum_{j=1}^{N-1} \left\{ \frac{1}{j(1-j/N) } \right\}^{2K} \quad,
\label{eq:Zapp}
\end{equation}
where $K=J/T$.
Three cases can be distinguished:
\begin{enumerate}
\item high temperatures, $K<1/2$: the sum in (\ref{eq:Zapp}) can be approximated by an integral, where $j/N$ is substituted 
by a continuous variable:
\begin{equation}
Z_{DW}^{*} = e^{-2(\gamma+1)K} N^{1-2K} \int_{0}^{1} dx \left\{ \frac{1}{x(1-x)} \right\}^{2K} \quad;
\label{eq:ZhT}
\end{equation}
in this case the DW free energy is given by
\begin{equation}
F_{DW}= -\frac{1-2K}{K }\ln N + O(N^0).
\label{eq:FDWappht}
\end{equation}
DW formation can proceed spontaneously;  the change in free energy is of order $-\ln N$. 
\item $K=1/2$: 
\begin{eqnarray}
\nonumber
Z_{DW}^{*}& = & e^{-(\gamma+1)} \sum_{j=1}^{N-1}  \frac{1}{j(1-j/N) }  \\
\nonumber
         & = & e^{-(\gamma+1)} \sum_{j=1}^{N-1} \left \{  \frac{1}{j}  + \frac{1}{N-j }\right \}\\
\nonumber
          &= & 2e^{-(\gamma+1)} \sum_{j=1}^{N-1}  \frac{1}{j}\\
         &=& 2e^{-(\gamma+1)} \ln N + O(1/N)  \quad.
\label{eq:Zm}
\end{eqnarray}
DW formation can still proceed spontaneously, with a free energy change of order $-\ln \ln N$. 
\item $K>1/2$: The leading term in the first half ($j<N/2) $ of the discrete sum (\ref{eq:Zapp}) - which is all that matters in the
thermodynamic limit - can be obtained by neglecting the second factor in the denominator. Since the terms are symmetric around
$j=N/2$, I estimate
\begin{equation}
Z_{DW}^{*} \approx e^{-2(\gamma+1)K} \>2 \zeta (2K)  \quad.
\label{eq:Zapplt}
\end{equation}
Note that in this case $Z_{DW}^{*}$  -and the DW free energy $-\ln Z_{DW}^{*}/K$ are of order $N^{0}$. 
Spontaneous DW formation occurs according to whether or not $Z_{DW}^{*}<1$. 
The critical temperature $T^{*}$ is obtained from the condition $Z_{DW}^{*}=1$, i.e. 
\begin{equation}
\frac{\ln2\zeta (2K^{*})}   { 2K^{*}}= 1+ \gamma \quad.
\label{eq:tcappr}
\end{equation} 
Numerical solution of (\ref{eq:tcappr}) gives a value $K^{*}=0.65136$. Note that the value is 
quite close to the Anderson-Yuwal \cite{AnYu} estimate $K_c = 0.635$. In fact, for 
$K^{*}$ not too far from $1/2$, it is possible to use the limiting form of the $\zeta$ function
$\zeta (2x) \approx 1/(2x-1)$; in this case (\ref{eq:tcappr}) becomes
\begin{equation}
-\frac{\ln (K^{*}-1/2)}  { 2K^{*}}= 1+ \gamma \quad.
\label{eq.TcYA}
\end{equation}
which is exactly the condition of \cite{AnYu}.  
\end{enumerate}

\subsection{The DW entropy: numerical calculations}
\begin{figure}
\vskip -.5truecm
\includegraphics[width=65mm]{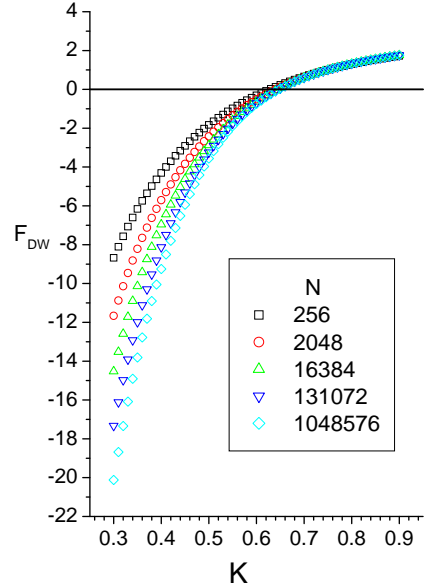}
\vskip -.5truecm
\caption{ 
{\small 
The numerically calculated free energy of a DW in units of $J$. as a function of the coupling constant $K$.
Note the different form of $N$-scaling for in the low and high $K$ regime (cf.  (\ref{eq:FDWappht}) and
(\ref{eq:Zapplt})). The dotted line marks the zero. 
}
}
\label{fig:DWfren}
\end{figure}

It is possible to use the exact expression (\ref{eq:DWenex}) of the DW energy  and calculate the partition function
 (\ref{eq:Zgen}) numerically. The resulting DW free energy as a function of $K$ for a range of chain sizes
is shown in Fig. \ref{fig:DWfren}.  Note the different form of $N$-scaling for in the low and 
high $K$ regimes, cf.  (\ref{eq:FDWappht}) and
(\ref{eq:Zapplt}). 
\begin{figure}
\vskip -.5truecm
\includegraphics[width=65mm]{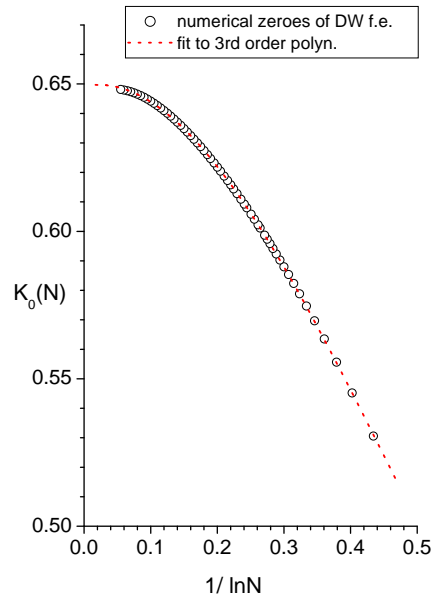}
\vskip -.5truecm
\caption{ 
{\small 
The zeroes of the numerically calculated DW free energy plotted as a function of 
$1/\ln N$. The continuous line is a fit of the type (\ref{eq:fit}) and yields $K^{*}=0.6492(2)$.
}
}
\label{fig:Kcnum}
\end{figure}

The critical point is defined by the zero of the DW excess free energy (intersection of 
each curve with the dotted line in Fig.  \ref{fig:DWfren}.) A third-order 
polynomial fit of zeroes obtained for sizes $10<N<8 \times 10^{7}$, i.e.
\begin{equation}
K_0(N) = K^{*} + \frac{a_{1}}{\ln N } +\frac{a_{2}}{(\ln N)^{2} }+\frac{a_{3}}{(\ln N)^{3} }
\label{eq:fit}
\end{equation}
(cf. Fig.   \ref{fig:Kcnum}) yields the estimate $K^{*}=0.6492(2)$ in the thermodynamic limit.

\section{Conclusions}
I have reviewed recent work on one-dimensional
lattice models with asymmetric on-site potentials, which are
known to exhibit thermodynamic instabilities, i.e. transitions from a bound to an unbound state,
at a temperature $T_c$.
Such nonlinear models generically admit exact, unbounded solutions which can be regarded as 
domain walls. Examination of DW thermodynamics shows that a DW becomes entropically 
stable (i.e. the energy cost of producing it is balanced by the entropic gain) at a temperature
$T^{*}$. In the
case treated analytically, $T^{*}=T_c$; in the case treated numerically, equality holds within
less than $1\%$.

A similar procedure was successfully applied to the one-dimensional ferromagnetic $S=1/2$ Ising
model with inverse square interactions. In that case, the DW thermodynamics leads to a $T^{*}$
which also differs by less than $1\%$ from the best available numerical estimate of the critical 
temperature.   

In summary, it appears that the vanishing of the DW free energy might be used
as a \lq\lq positive\rq\rq criterion for the occurrence of a phase transition in one-dimensional
systems, and at the same time provide an estimate of the critical temperature. In this sense, one might be
tempted to conclude that phase transitions in one dimensional systems
are indeed driven by DW formation. 
\par

\end{document}